\documentclass[twocolumn, traditabstract, longauth]{aa}

\usepackage{amsmath}
\usepackage{natbib}
\usepackage{graphicx}
\usepackage[varg]{txfonts}
\usepackage{color}
\usepackage{url}
\usepackage{hyperref}
\usepackage{breakurl}

\makeatletter
\newcommand\tinyv{\@setfontsize\tinyv{4pt}{6}}
\renewcommand*{\@fnsymbol}[1]{\ifcase#1\or*\or$\dagger$\or$\ddagger$\or**\or$\dagger\dagger$\or$\ddagger\ddagger$\fi}
\makeatother

\newcommand{\gray}{$\gamma$-ray}
\newcommand{\grays}{$\gamma$ rays}

\def\apjl{ApJ}                % Astrophysical Journal, Letters
\def\apjs{ApJS}               % Astrophysical Journal, Supplement
\def\aap{A\&A}                % Astronomy and Astrophysics
\makeatletter
\renewcommand*{\@fnsymbol}[1]{\ifcase#1\or*\or$\dagger$\or$\ddagger$\or**\or$\dagger\dagger$\or$\ddagger\ddagger$\fi}
\makeatother

\begin{document}

\title{The $\gamma$-ray spectrum of the core of Centaurus A as observed with H.E.S.S. and \textit{Fermi}-LAT}

\authorrunning{H.E.S.S and {Fermi}-LAT Collaborations}
\titlerunning{Spectrum of the $\gamma$-ray core of Centaurus A}

\author{\fontsize{10pt}{12pt}\selectfont
H.E.S.S. Collaboration \and H.~Abdalla \inst{\ref{NWU}} \and
A.~Abramowski \inst{\ref{HH}} \and F.~Aharonian
\inst{\ref{MPIK},\ref{DIAS},\ref{NASRA}} \and F.~Ait~Benkhali
\inst{\ref{MPIK}} \and E.O.~Ang\"uner \inst{\ref{IFJPAN}} \and
M.~Arakawa \inst{\ref{Rikkyo}} \and C.~Armand \inst{\ref{LAPP}} \and
M.~Arrieta \inst{\ref{LUTH}} \and M.~Backes \inst{\ref{UNAM}} \and
A.~Balzer \inst{\ref{GRAPPA}} \and M.~Barnard \inst{\ref{NWU}} \and
Y.~Becherini\protect\footnotemark[1] \inst{\ref{Linnaeus}} \and
J.~Becker~Tjus \inst{\ref{RUB}} \and D.~Berge \inst{\ref{DESY}} \and
S.~Bernhard \inst{\ref{LFUI}} \and K.~Bernl\"ohr \inst{\ref{MPIK}}
\and R.~Blackwell \inst{\ref{Adelaide}} \and M.~B\"ottcher
\inst{\ref{NWU}} \and C.~Boisson \inst{\ref{LUTH}} \and J.~Bolmont
\inst{\ref{LPNHE}} \and S.~Bonnefoy \inst{\ref{DESY}} \and P.~Bordas
\inst{\ref{MPIK}} \and J.~Bregeon \inst{\ref{LUPM}} \and F.~Brun
\inst{\ref{CENB}} \and P.~Brun \inst{\ref{IRFU}} \and M.~Bryan
\inst{\ref{GRAPPA}} \and M.~B\"{u}chele \inst{\ref{ECAP}} \and
T.~Bulik \inst{\ref{UWarsaw}} \and M.~Capasso \inst{\ref{IAAT}} \and
S.~Caroff \inst{\ref{LLR}} \and A.~Carosi \inst{\ref{LAPP}} \and
S.~Casanova \inst{\ref{IFJPAN},\ref{MPIK}} \and M.~Cerruti
\inst{\ref{LPNHE}} \and N.~Chakraborty \inst{\ref{MPIK}} \and
R.C.G.~Chaves \inst{\ref{LUPM},\ref{CurieChaves}} \and A.~Chen
\inst{\ref{WITS}} \and J.~Chevalier \inst{\ref{LAPP}} \and
S.~Colafrancesco \inst{\ref{WITS}} \and B.~Condon \inst{\ref{CENB}}
\and J.~Conrad \inst{\ref{OKC},\ref{FellowConrad}} \and I.D.~Davids
\inst{\ref{UNAM}} \and J.~Decock \inst{\ref{IRFU}} \and C.~Deil
\inst{\ref{MPIK}} \and J.~Devin \inst{\ref{LUPM}} \and P.~deWilt
\inst{\ref{Adelaide}} \and L.~Dirson \inst{\ref{HH}} \and
A.~Djannati-Ata\"i \inst{\ref{APC}} \and A.~Donath \inst{\ref{MPIK}}
\and L.O'C.~Drury \inst{\ref{DIAS}} \and J.~Dyks \inst{\ref{NCAC}}
\and T.~Edwards \inst{\ref{MPIK}} \and K.~Egberts \inst{\ref{UP}}
\and G.~Emery \inst{\ref{LPNHE}} \and J.-P.~Ernenwein
\inst{\ref{CPPM}} \and S.~Eschbach \inst{\ref{ECAP}} \and C.~Farnier
\inst{\ref{OKC},\ref{Linnaeus}} \and S.~Fegan \inst{\ref{LLR}} \and
M.V.~Fernandes \inst{\ref{HH}} \and A.~Fiasson \inst{\ref{LAPP}}
\and G.~Fontaine \inst{\ref{LLR}} \and S.~Funk \inst{\ref{ECAP}}
\and M.~F\"u{\ss}ling \inst{\ref{DESY}} \and S.~Gabici
\inst{\ref{APC}} \and Y.A.~Gallant \inst{\ref{LUPM}} \and
T.~Garrigoux \inst{\ref{NWU}} \and F.~Gat{\'e} \inst{\ref{LAPP}}
\and G.~Giavitto \inst{\ref{DESY}} \and D.~Glawion \inst{\ref{LSW}}
\and J.F.~Glicenstein \inst{\ref{IRFU}} \and D.~Gottschall
\inst{\ref{IAAT}} \and M.-H.~Grondin \inst{\ref{CENB}} \and J.~Hahn
\inst{\ref{MPIK}} \and M.~Haupt \inst{\ref{DESY}} \and J.~Hawkes
\inst{\ref{Adelaide}} \and G.~Heinzelmann \inst{\ref{HH}} \and
G.~Henri \inst{\ref{Grenoble}} \and G.~Hermann \inst{\ref{MPIK}}
\and J.A.~Hinton \inst{\ref{MPIK}} \and W.~Hofmann \inst{\ref{MPIK}}
\and C.~Hoischen \inst{\ref{UP}} \and T.~L.~Holch \inst{\ref{HUB}}
\and M.~Holler \inst{\ref{LFUI}} \and D.~Horns \inst{\ref{HH}} \and
A.~Ivascenko \inst{\ref{NWU}} \and H.~Iwasaki \inst{\ref{Rikkyo}}
\and A.~Jacholkowska \inst{\ref{LPNHE}} \and M.~Jamrozy
\inst{\ref{UJK}} \and D.~Jankowsky \inst{\ref{ECAP}} \and
F.~Jankowsky \inst{\ref{LSW}} \and M.~Jingo \inst{\ref{WITS}} \and
L.~Jouvin \inst{\ref{APC}} \and I.~Jung-Richardt \inst{\ref{ECAP}}
\and M.A.~Kastendieck \inst{\ref{HH}} \and K.~Katarzy{\'n}ski
\inst{\ref{NCUT}} \and M.~Katsuragawa \inst{\ref{JAXA}} \and U.~Katz
\inst{\ref{ECAP}} \and D.~Kerszberg \inst{\ref{LPNHE}} \and
D.~Khangulyan \inst{\ref{Rikkyo}} \and B.~Kh\'elifi \inst{\ref{APC}}
\and J.~King \inst{\ref{MPIK}} \and S.~Klepser \inst{\ref{DESY}}
\and D.~Klochkov \inst{\ref{IAAT}} \and W.~Klu\'{z}niak
\inst{\ref{NCAC}} \and Nu.~Komin \inst{\ref{WITS}} \and K.~Kosack
\inst{\ref{IRFU}} \and S.~Krakau \inst{\ref{RUB}} \and M.~Kraus
\inst{\ref{ECAP}} \and P.P.~Kr\"uger \inst{\ref{NWU}} \and H.~Laffon
\inst{\ref{CENB}} \and G.~Lamanna \inst{\ref{LAPP}} \and J.~Lau
\inst{\ref{Adelaide}} \and J.~Lefaucheur \inst{\ref{LUTH}} \and
A.~Lemi\`ere \inst{\ref{APC}} \and M.~Lemoine-Goumard
\inst{\ref{CENB}} \and J.-P.~Lenain \inst{\ref{LPNHE}} \and E.~Leser
\inst{\ref{UP}} \and T.~Lohse \inst{\ref{HUB}} \and M.~Lorentz
\inst{\ref{IRFU}} \and R.~Liu \inst{\ref{MPIK}} \and R.~L\'opez-Coto
\inst{\ref{MPIK}} \and I.~Lypova \inst{\ref{DESY}} \and D.~Malyshev
\inst{\ref{IAAT}} \and V.~Marandon \inst{\ref{MPIK}} \and
A.~Marcowith \inst{\ref{LUPM}} \and C.~Mariaud \inst{\ref{LLR}} \and
R.~Marx \inst{\ref{MPIK}} \and G.~Maurin \inst{\ref{LAPP}} \and
N.~Maxted \inst{\ref{Adelaide},\ref{MaxtedNowAt}} \and M.~Mayer
\inst{\ref{HUB}} \and P.J.~Meintjes \inst{\ref{UFS}} \and M.~Meyer
\inst{\ref{OKC},\ref{MeyerNowAt}} \and A.M.W.~Mitchell
\inst{\ref{MPIK}} \and R.~Moderski \inst{\ref{NCAC}} \and M.~Mohamed
\inst{\ref{LSW}} \and L.~Mohrmann \inst{\ref{ECAP}} \and K.~Mor{\aa}
\inst{\ref{OKC}} \and E.~Moulin \inst{\ref{IRFU}} \and T.~Murach
\inst{\ref{DESY}} \and S.~Nakashima  \inst{\ref{JAXA}} \and
M.~de~Naurois \inst{\ref{LLR}} \and H.~Ndiyavala  \inst{\ref{NWU}}
\and F.~Niederwanger \inst{\ref{LFUI}} \and J.~Niemiec
\inst{\ref{IFJPAN}} \and L.~Oakes \inst{\ref{HUB}} \and P.~O'Brien
\inst{\ref{Leicester}} \and H.~Odaka \inst{\ref{JAXA}} \and S.~Ohm
\inst{\ref{DESY}} \and M.~Ostrowski \inst{\ref{UJK}} \and I.~Oya
\inst{\ref{DESY}} \and M.~Padovani \inst{\ref{LUPM}} \and M.~Panter
\inst{\ref{MPIK}} \and R.D.~Parsons \inst{\ref{MPIK}} \and
N.W.~Pekeur \inst{\ref{NWU}} \and G.~Pelletier \inst{\ref{Grenoble}}
\and C.~Perennes \inst{\ref{LPNHE}} \and P.-O.~Petrucci
\inst{\ref{Grenoble}} \and B.~Peyaud \inst{\ref{IRFU}} \and Q.~Piel
\inst{\ref{LAPP}} \and S.~Pita \inst{\ref{APC}} \and V.~Poireau
\inst{\ref{LAPP}} \and D.A.~Prokhorov\protect\footnotemark[1]
\inst{\ref{Linnaeus}, \ref{WITS}} \and H.~Prokoph
\inst{\ref{GRAPPAHE}} \and G.~P\"uhlhofer \inst{\ref{IAAT}} \and
M.~Punch \inst{\ref{APC},\ref{Linnaeus}} \and A.~Quirrenbach
\inst{\ref{LSW}} \and S.~Raab \inst{\ref{ECAP}} \and R.~Rauth
\inst{\ref{LFUI}} \and A.~Reimer \inst{\ref{LFUI}} \and O.~Reimer
\inst{\ref{LFUI}} \and M.~Renaud \inst{\ref{LUPM}} \and
R.~de~los~Reyes \inst{\ref{MPIK}} \and
F.~Rieger\protect\footnotemark[1]
\inst{\ref{MPIK},\ref{FellowRieger}} \and L.~Rinchiuso
\inst{\ref{IRFU}} \and C.~Romoli \inst{\ref{DIAS}} \and G.~Rowell
\inst{\ref{Adelaide}} \and B.~Rudak \inst{\ref{NCAC}} \and
C.B.~Rulten \inst{\ref{LUTH}} \and V.~Sahakian
\inst{\ref{YPI},\ref{NASRA}} \and S.~Saito \inst{\ref{Rikkyo}} \and
D.A.~Sanchez \inst{\ref{LAPP}} \and A.~Santangelo \inst{\ref{IAAT}}
\and M.~Sasaki \inst{\ref{ECAP}} \and R.~Schlickeiser
\inst{\ref{RUB}} \and F.~Sch\"ussler \inst{\ref{IRFU}} \and
A.~Schulz \inst{\ref{DESY}} \and U.~Schwanke \inst{\ref{HUB}} \and
S.~Schwemmer \inst{\ref{LSW}} \and M.~Seglar-Arroyo
\inst{\ref{IRFU}} \and A.S.~Seyffert \inst{\ref{NWU}} \and N.~Shafi
\inst{\ref{WITS}} \and I.~Shilon \inst{\ref{ECAP}} \and
K.~Shiningayamwe \inst{\ref{UNAM}} \and R.~Simoni
\inst{\ref{GRAPPA}} \and H.~Sol \inst{\ref{LUTH}} \and F.~Spanier
\inst{\ref{NWU}} \and M.~Spir-Jacob \inst{\ref{APC}} \and
{\L.}~Stawarz \inst{\ref{UJK}} \and R.~Steenkamp \inst{\ref{UNAM}}
\and C.~Stegmann \inst{\ref{UP},\ref{DESY}} \and C.~Steppa
\inst{\ref{UP}} \and I.~Sushch \inst{\ref{NWU}} \and T.~Takahashi
\inst{\ref{JAXA}} \and J.-P.~Tavernet \inst{\ref{LPNHE}} \and
T.~Tavernier \inst{\ref{IRFU}} \and A.M.~Taylor \inst{\ref{DESY}}
\and R.~Terrier \inst{\ref{APC}} \and L.~Tibaldo \inst{\ref{MPIK}}
\and D.~Tiziani \inst{\ref{ECAP}} \and M.~Tluczykont \inst{\ref{HH}}
\and C.~Trichard \inst{\ref{CPPM}} \and M.~Tsirou \inst{\ref{LUPM}}
\and N.~Tsuji \inst{\ref{Rikkyo}} \and R.~Tuffs \inst{\ref{MPIK}}
\and Y.~Uchiyama \inst{\ref{Rikkyo}} \and D.J.~van~der~Walt
\inst{\ref{NWU}} \and C.~van~Eldik \inst{\ref{ECAP}} \and
C.~van~Rensburg \inst{\ref{NWU}} \and B.~van~Soelen \inst{\ref{UFS}}
\and G.~Vasileiadis \inst{\ref{LUPM}} \and J.~Veh \inst{\ref{ECAP}}
\and C.~Venter \inst{\ref{NWU}} \and A.~Viana
\inst{\ref{MPIK},\ref{VianaNowAt}} \and P.~Vincent
\inst{\ref{LPNHE}} \and J.~Vink \inst{\ref{GRAPPA}} \and F.~Voisin
\inst{\ref{Adelaide}} \and H.J.~V\"olk \inst{\ref{MPIK}} \and
T.~Vuillaume \inst{\ref{LAPP}} \and Z.~Wadiasingh \inst{\ref{NWU}}
\and S.J.~Wagner \inst{\ref{LSW}} \and P.~Wagner \inst{\ref{HUB}}
\and R.M.~Wagner \inst{\ref{OKC}} \and R.~White \inst{\ref{MPIK}}
\and A.~Wierzcholska \inst{\ref{IFJPAN}} \and P.~Willmann
\inst{\ref{ECAP}} \and A.~W\"ornlein \inst{\ref{ECAP}} \and
D.~Wouters \inst{\ref{IRFU}} \and R.~Yang \inst{\ref{MPIK}} \and
D.~Zaborov \inst{\ref{LLR}} \and M.~Zacharias \inst{\ref{NWU}} \and
R.~Zanin \inst{\ref{MPIK}} \and A.A.~Zdziarski \inst{\ref{NCAC}}
\and A.~Zech \inst{\ref{LUTH}} \and F.~Zefi \inst{\ref{LLR}} \and
A.~Ziegler \inst{\ref{ECAP}} \and J.~Zorn \inst{\ref{MPIK}} \and
N.~\.Zywucka \inst{\ref{UJK}} \\
\textit{Fermi}-LAT collaboration \and
J.~D.~Magill\protect\footnotemark[1] \inst{\ref{Maryland}} \and
S.~Buson \inst{\ref{NASA}, \ref{NASAfellow}} \and C.~C.~Cheung
\inst{\ref{Naval}} \and J.~S.~Perkins \inst{\ref{NASA}} \and
Y.~Tanaka\inst{\ref{Hiroshima}} }

\institute{Centre for Space Research, North-West University,
Potchefstroom 2520, South Africa \label{NWU} \and Universit\"at
Hamburg, Institut f\"ur Experimentalphysik, Luruper Chaussee 149, D
22761 Hamburg, Germany \label{HH} \and Max-Planck-Institut f\"ur
Kernphysik, P.O. Box 103980, D 69029 Heidelberg, Germany
\label{MPIK} \and Dublin Institute for Advanced Studies, 31
Fitzwilliam Place, Dublin 2, Ireland \label{DIAS} \and National
Academy of Sciences of the Republic of Armenia,  Marshall Baghramian
Avenue, 24, 0019 Yerevan, Republic of Armenia \label{NASRA} \and
Instytut Fizyki J\c{a}drowej PAN, ul. Radzikowskiego 152, 31-342
Krak{\'o}w, Poland \label{IFJPAN} \and Department of Physics, Rikkyo
University, 3-34-1 Nishi-Ikebukuro, Toshima-ku, Tokyo 171-8501,
Japan \label{Rikkyo} \and Laboratoire d'Annecy de Physique des
Particules, Universit\'{e} Savoie Mont-Blanc, CNRS/IN2P3, F-74941
Annecy-le-Vieux, France \label{LAPP} \and LUTH, Observatoire de
Paris, PSL Research University, CNRS, Universit\'e Paris Diderot, 5
Place Jules Janssen, 92190 Meudon, France \label{LUTH} \and
University of Namibia, Department of Physics, Private Bag 13301,
Windhoek, Namibia \label{UNAM} \and GRAPPA, Anton Pannekoek
Institute for Astronomy, University of Amsterdam,  Science Park 904,
1098 XH Amsterdam, The Netherlands \label{GRAPPA} \and Department of
Physics and Electrical Engineering, Linnaeus University,  351 95
V\"axj\"o, Sweden \label{Linnaeus} \and Institut f\"ur Theoretische
Physik, Lehrstuhl IV: Weltraum und Astrophysik, Ruhr-Universit\"at
Bochum, D 44780 Bochum, Germany \label{RUB} \and DESY, D-15738
Zeuthen, Germany \label{DESY} \and Institut f\"ur Astro- und
Teilchenphysik, Leopold-Franzens-Universit\"at Innsbruck, A-6020
Innsbruck, Austria \label{LFUI} \and School of Physical Sciences,
University of Adelaide, Adelaide 5005, Australia \label{Adelaide}
\and Sorbonne Universit\'es, UPMC Universit\'e Paris 06,
Universit\'e Paris Diderot, Sorbonne Paris Cit\'e, CNRS, Laboratoire
de Physique Nucl\'eaire et de Hautes Energies (LPNHE), 4 place
Jussieu, F-75252, Paris Cedex 5, France \label{LPNHE} \and
Laboratoire Univers et Particules de Montpellier, Universit\'e
Montpellier, CNRS/IN2P3,  CC 72, Place Eug\`ene Bataillon, F-34095
Montpellier Cedex 5, France \label{LUPM} \and Universit\'e Bordeaux,
CNRS/IN2P3, Centre d'\'Etudes Nucl\'eaires de Bordeaux Gradignan,
33175 Gradignan, France \label{CENB} \and IRFU, CEA, Universit\'e
Paris-Saclay, F-91191 Gif-sur-Yvette, France \label{IRFU} \and
Friedrich-Alexander-Universit\"at Erlangen-N\"urnberg, Erlangen
Centre for Astroparticle Physics, Erwin-Rommel-Str. 1, D 91058
Erlangen, Germany \label{ECAP} \and Astronomical Observatory, The
University of Warsaw, Al. Ujazdowskie 4, 00-478 Warsaw, Poland
\label{UWarsaw} \and Institut f\"ur Astronomie und Astrophysik,
Universit\"at T\"ubingen, Sand 1, D 72076 T\"ubingen, Germany
\label{IAAT} \and Laboratoire Leprince-Ringuet, Ecole Polytechnique,
CNRS/IN2P3, F-91128 Palaiseau, France \label{LLR} \and Funded by EU
FP7 Marie Curie, grant agreement No. PIEF-GA-2012-332350
\label{CurieChaves} \and School of Physics, University of the
Witwatersrand, 1 Jan Smuts Avenue, Braamfontein, Johannesburg, 2050
South Africa \label{WITS} \and Oskar Klein Centre, Department of
Physics, Stockholm University, Albanova University Center, SE-10691
Stockholm, Sweden \label{OKC} \and Wallenberg Academy Fellow
\label{FellowConrad} \and APC, AstroParticule et Cosmologie,
Universit\'{e} Paris Diderot, CNRS/IN2P3, CEA/Irfu, Observatoire de
Paris, Sorbonne Paris Cit\'{e}, 10, rue Alice Domon et L\'{e}onie
Duquet, 75205 Paris Cedex 13, France \label{APC} \and Nicolaus
Copernicus Astronomical Center, Polish Academy of Sciences, ul.
Bartycka 18, 00-716 Warsaw, Poland \label{NCAC} \and Institut f\"ur
Physik und Astronomie, Universit\"at Potsdam,
Karl-Liebknecht-Strasse 24/25, D 14476 Potsdam, Germany \label{UP}
\and Aix Marseille Universit\'e, CNRS/IN2P3, CPPM, Marseille, France
\label{CPPM} \and Landessternwarte, Universit\"at Heidelberg,
K\"onigstuhl, D 69117 Heidelberg, Germany \label{LSW} \and Univ.
Grenoble Alpes, CNRS, IPAG, F-38000 Grenoble, France
\label{Grenoble}  \and Institut f\"ur Physik, Humboldt-Universit\"at
zu Berlin, Newtonstr. 15, D 12489 Berlin, Germany \label{HUB} \and
Obserwatorium Astronomiczne, Uniwersytet Jagiello{\'n}ski, ul. Orla
171, 30-244 Krak{\'o}w, Poland \label{UJK} \and Centre for
Astronomy, Faculty of Physics, Astronomy and Informatics, Nicolaus
Copernicus University, Grudziadzka 5, 87-100 Torun, Poland
\label{NCUT} \and Japan Aerpspace Exploration Agency (JAXA),
Institute of Space and Astronautical Science (ISAS), 3-1-1
Yoshinodai, Chuo-ku, Sagamihara, Kanagawa 229-8510,  Japan
\label{JAXA} \and Department of Physics, University of the Free
State, PO Box 339, Bloemfontein 9300, South Africa \label{UFS} \and
Now at Kavli Institute for Particle Astrophysics and Cosmology,
Department of Physics and SLAC National Accelerator Laboratory,
Stanford University, Stanford, California 94305, USA
\label{MeyerNowAt} \and Now at The School of Physics, The University
of New South Wales, Sydney, 2052, Australia \label{MaxtedNowAt} \and
Department of Physics and Astronomy, The University of Leicester,
University Road, Leicester, LE1 7RH, United Kingdom
\label{Leicester} \and GRAPPA, Anton Pannekoek Institute for
Astronomy and Institute of High-Energy Physics, University of
Amsterdam,  Science Park 904, 1098 XH Amsterdam, The Netherlands
\label{GRAPPAHE} \and Heisenberg Fellow (DFG), ITA Universit\"at
Heidelberg, Germany \label{FellowRieger} \and Yerevan Physics
Institute, 2 Alikhanian Brothers St., 375036 Yerevan, Armenia
\label{YPI} \and Now at Instituto de F\'{i}sica de S\~{a}o Carlos,
Universidade de S\~{a}o Paulo, Av. Trabalhador S\~{a}o-carlense, 400
- CEP 13566-590, S\~{a}o Carlos, SP, Brazil \label{VianaNowAt} \and
Department of Physics and Department of Astronomy, University of
Maryland, College Park, MD 20742, USA \label{Maryland} \and NASA
Goddard Space Flight Center, Greenbelt, MD 20771, USA \label{NASA}
\and NASA Postdoctoral Program Fellow, USA \label{NASAfellow} \and
Space Science Division, Naval Research Laboratory, Washington, DC
20375-5352, USA \label{Naval} \and Hiroshima Astrophysical Science
Center, Hiroshima University, Higashi-Hiroshima, Hiroshima 739-8526,
Japan \label{Hiroshima}
%% Affiliation of people who left the collaboration
}

\offprints{H.E.S.S. and \textit{Fermi}-LAT collaborations,
\protect\\\email{\href{mailto:contact.hess@hess-experiment.eu}{contact.hess@hess-experiment.eu};
jmagill@umd.edu}
%\protect\\\protect\footnotemark[1] Corresponding authors
%%\protect\\\protect\footnotemark[2] Deceased
}

%\date{Received 21 November 2017}

\abstract{Centaurus A (Cen A) is the nearest radio galaxy discovered
as a very-high-energy (VHE; 100 GeV-100 TeV) \gray{} source by the
High Energy Stereoscopic System (H.E.S.S.).
It is a faint VHE \gray{} emitter, though its
VHE flux exceeds both the extrapolation from early {\it Fermi}-LAT
observations as well as expectations from a (misaligned) single-zone
synchrotron-self Compton (SSC) description. The latter
satisfactorily reproduces the emission from Cen~A at lower energies
up to a few GeV. New observations with H.E.S.S., comparable in
exposure time to those previously reported, were performed and eight
years of {\it Fermi}-LAT data were accumulated to clarify the
spectral characteristics of the \gray{} emission from the core of
Cen~A. The results allow us for the first time to achieve
the goal of constructing a representative, contemporaneous
$\gamma$-ray core spectrum of Cen~A over almost five orders of
magnitude in energy. Advanced analysis methods, including the
template fitting method, allow detection in
the VHE range of the core with a statistical significance of
12$\sigma$ on the basis of 213 hours of total exposure time. The
spectrum in the energy range of 250 GeV-6 TeV is compatible with a
power-law function with a photon index
$\Gamma=2.52\pm0.13_{\mathrm{stat}}\pm0.20_{\mathrm{sys}}$. An
updated {\it Fermi}-LAT analysis provides evidence for spectral
hardening by $\Delta\Gamma\simeq0.4\pm0.1$ at $\gamma$-ray energies
above $2.8\protect\substack{+1.0
\\ -0.6}$\,GeV at a level of $4.0\sigma$. The fact that the spectrum
hardens at GeV energies and extends into
the VHE regime disfavour a single-zone SSC interpretation for the
overall spectral energy distribution (SED) of the core and is
suggestive of a new $\gamma$-ray emitting component connecting the
high-energy emission above the break energy to the one observed at
VHE energies. The absence of significant variability at both GeV and
TeV energies does not yet allow disentanglement of the physical
nature of this component, though a jet-related origin is possible
and a simple two-zone SED model fit is provided to this end.}


\maketitle

%% Redefine numeric symbols for footnotes
\makeatletter
\renewcommand*{\@fnsymbol}[1]{\ifcase#1\@arabic{#1}\fi}
\makeatother

\section{Introduction}
Active galaxies host a small, bright core of non-thermal emission.
At a distance of $d\simeq3.8$ Mpc, Centaurus A (Cen A) is the
nearest active galaxy \citep[][]{israel98, harris10}. Its proximity
has allowed for a detailed morphological analysis over angular
scales ranging from milli-arcseconds to several degrees
(1$^{\circ}\simeq65$ kpc). A variety of structures powered by its
active galactic nucleus (AGN) have been discovered using
observations in radio \citep[e.g.][]{hardcastle03, horiuchi06,
muller14}, infrared \citep[e.g.][]{brookes06, hardcastle06,
meisenheimer07}, X-ray \citep[e.g.][]{kraft02,hardcastle03}, and
\gray{} \citep[e.g.][]{abdo10, abdo10i, yang12} bands. These
structures include a radio emitting core with a size of $\leq
10^{-2}$ pc, a parsec-scale jet
and counter-jet system, a kiloparsec-scale jet
and inner lobes, up to giant outer lobes with a length of hundreds
of kiloparsecs.

Based on its radio properties, Cen A has been classified as a radio
galaxy of Fanaroff-Riley type I \citep[][]{fanaroff74}. According to
AGN unification schemes, radio galaxies of this type are thought to
correspond to BL Lacertae (BL Lac) objects viewed from the side, the latter
showing jets aligned along the line of sight and corresponding to a subclass
of blazars \citep[][]{urry95}. BL Lac objects are the most abundant
class of known extragalactic very-high-energy (VHE)
emitters\footnote{\url{http://tevcat.uchicago.edu/}.}, and exhibit
double-peaked spectral energy distributions (SEDs). It is commonly
thought that their low-frequency emission in the radio to ultraviolet
(and X-ray, for high-peaked BL Lacs) band is synchrotron emission from
relativistic electrons within a blob (zone) moving at relativistic
speeds in the jet. Synchrotron self-absorption implies that
the lower-frequency observed radio emission cannot be produced by a
compact blob, and is likely produced by synchrotron from a larger jet
component. The high-energy emission (hard X-ray to VHE \gray{}) from
high-peaked BL Lac type objects has been satisfactorily modelled as
synchrotron self-Compton (SSC) radiation resulting from the inverse
Compton upscattering of synchrotron photons by the same relativistic
electrons that produced the synchrotron radiation
\citep[][]{maraschi92, bloom96}, although other more complex models
(involving e.g. external inverse Compton emission, hadronic
interactions, or multiple zones) are conceivable \citep[][]{Reimer13}.

At a few tens of keV to GeV photon energies, Cen A was detected by
all instruments on board the \textit{Compton} Gamma-Ray Observatory
(BATSE, OSSE, COMPTEL, and EGRET; the acronyms are
described in Appendix B.)
in the period 1991-1995 revealing
a high-energy peak in the SED at an energy of $\sim0.1$ MeV
\citep[see][]{kinzer95, steinle98, sreekumar99}. An earlier
investigation found that it is possible to fit the data ranging from
the radio band to the \gray{} band using a single-zone SSC model
\citep[][]{chiaberge01}, but this implies a low flux at VHE. High-energy
and VHE \gray{} observations are thus important to test the validity
of the SSC scenario for modelling of the SED of radio galaxies.

The discovery of Cen A as an emitter of VHE \grays{} was reported on
the basis of 115 hr of observation (labelled data set A in this
study) with the High Energy Stereoscopic System (H.E.S.S.) performed
from April 2004 to July 2008 \citep[][]{paper1}. The signal from the
region containing the radio core, the
parsec-scale jet, and the kiloparsec-scale
jet was detected with a statistical significance of
5.0$\sigma$. In this paper, we refer to this region as the Cen A
\gray{} core. Subsequent survey observations at high energies (HE;
100 MeV - 100 GeV) were performed by the Large Area Telescope (LAT)
on board the \textit{Fermi Gamma Ray Space Telescope} (FGST)
launched in June 2008 \citep[][]{atwood09}. During the first three
months of science operation, started on August 4, 2008,
\textit{Fermi}-LAT confirmed the EGRET detection of the Cen A
\gray{} core \citep[][]{BrightList2009}. Spectral analysis and
modelling based on ten months of \textit{Fermi}-LAT observations
\citep[][]{abdo10} suggested the high-energy \gray{} emission up to
$\sim10$ GeV to be compatible with a single power law, yet indicated
that a single-zone SSC model would be unable to account for the
(non-contemporaneous) higher energy TeV emission observed by
H.E.S.S. in 2004-2008. The analysis of extended \textit{Fermi}-LAT
data sets has in the meantime provided increasing evidence for a
substantial spectral break above a few GeV \citep[][]{sahakyan13,
brown16}. This supports the conclusion that the TeV emission
observed in 2004-2008 with H.E.S.S. belongs to a distinct, separate
spectral component.

In this paper, we present the results of long-term observations of
the Cen A \gray{} core performed both with H.E.S.S. and with
\textit{Fermi}-LAT. These include new (more than 100 hr) VHE
observations of the Cen A \gray{} core with H.E.S.S. (data set B)
performed when the FGST was already in orbit. We report results of
the spectral analysis of the complete H.E.S.S. data set (Sect.~2)
with an exposure time that is twice that used in the previously
published data set A, as well as an update (Sect.~3) of the spectrum
of the Cen A \gray{} core obtained with \textit{Fermi}-LAT at GeV
energies. The results are discussed and put into wider context in
Section~4.

\section{H.E.S.S. observations and results}
Cen A is a weak VHE source with a measured integral flux above 250
GeV of about $0.8\%$ of the flux of the Crab Nebula.\footnote{The
observed integral flux of the Crab Nebula above 1 TeV is
$(2.26\pm0.08_{\mathrm{stat}}\pm0.45_{\mathrm{sys}})\times10^{-11}$
cm$^{-2}$s$^{-1}$ \citep[][]{crab}.} The discovery of faint VHE
\gray{} emission from Cen A motivated further observations with
H.E.S.S., which were performed in 2009-2010. In this section, we
report the results of the Cen A observations with H.E.S.S. taken
between 2004 and 2010. It includes a re-analysis of the H.E.S.S.
data taken between 2004 and 2008 using refined methods. Using the
combined H.E.S.S. data set (data sets A+B), we perform a detailed
study of the VHE spectrum of Cen A.

\subsection{Observations and analysis}

The H.E.S.S. experiment is an array of five imaging atmospheric
Cherenkov telescopes\footnote{The fifth telescope with its
28-metre-sized mirror was added to the array in summer 2012 during
the H.E.S.S. phase-II upgrade, lowering the energy threshold of the
array.} located in the southern hemisphere in Namibia
($23^{\circ}16^{\prime}18^{\prime\prime}$ S, $16^{\circ}30^{\prime}
00^{\prime\prime}$ E) at an altitude of 1800 m above sea level
\citep[][]{crab}. At the time of the observations used in this
paper, the H.E.S.S. array consisted of four 12-metre telescopes. The
telescopes, arranged in a square with 120-m sides, have been in
operation since 2004 \citep[][]{hinton2004}. Each of these
telescopes covers a field of view of 5$^{\circ}$ diameter. H.E.S.S.
employs the stereoscopic imaging atmospheric Cherenkov technique
\citep[e.g.][]{daum97} and is sensitive with these telescopes to
\grays{} above an energy threshold of $\sim0.1$ TeV for observations
at zenith, up to energies of tens of TeV. The energy
threshold increases with zenith angle.
The observations of Cen A with H.E.S.S.
reported in this paper were performed in wobble mode, that is with the
target typically offset by about 0.5$^{\circ}$ or 0.7$^{\circ}$ from
the pointing direction, allowing simultaneous background estimation
in the same field of view \citep[][]{berge07}. The data were
recorded in 28-minute exposures, called runs, which are chosen to
minimise systematic changes in instrumental response. The
observations of Cen A were carried out during the January-July
visibility window.

Data set A was taken between April 2004 and July 2008, and 111 hr of
good-quality data \citep[following a cut on the satisfactory
hardware state of the cameras and good atmospheric conditions, as
described in][]{crab} were recorded during 261 runs. The mean zenith
angle of these observations is $24^{\circ}$. The results of a
re-analysis of the data set A are presented in Section
\ref{sect221}.
The new data set B was taken from 2009 to 2010 and consists of 241
runs corresponding to 102 hr of additional exposure. The mean zenith
angle of these observations is $23^{\circ}$. The total exposure time
(data set A and B) adds up to 213 hr.
Data set A was taken prior to the launch of the FGST,
while the new data set B presented here was taken after the launch
of the FGST. The consistency between the results of the H.E.S.S.
observations of Cen A in these two time intervals, that is, the lack of
flux variability along with no change in spectral parameters, is of
importance to substantiate a simultaneous spectral fit of both the
HE
and VHE data. The results of an analysis of data set B and of a
joint analysis of the two data sets are presented in Sections
\ref{sect222} and \ref{sect223}, respectively.

The Image Pixel-wise fit for Atmospheric Cherenkov Telescope
(ImPACT) analysis \citep[][]{Parsons2014} was used to process the
H.E.S.S. data. The gain of the ImPACT analysis in sensitivity is of
more than a factor of 1.5 over traditional image moments-fitting
(Hillas-based) analyses, used by \citet[][]{paper1}.

The \texttt{std\_ImPACT} cut configuration, which requires a minimum
of 60 photo-electrons per image, was used. The On-source counts were
taken from the circular region around the Cen A radio core. The same
On-region was selected for analyses of the data sets A and B, and
for a combined analysis.
The reflected-region background method with multiple Off-source
regions was used for spectral measurements. Given the angular
resolution of H.E.S.S., the giant outer lobes are expected to
negligibly affect the VHE results.
Thus, the results of the H.E.S.S. data analysis for Cen A reported
here are based on twice the exposure and a more sensitive analysis
of data set A than that used in the publication in 2009. To
cross-check the results, an independent analysis method based on a
multivariate combination of discriminant variables using the
physical shower properties \citep[][]{Becherini2011} has been
applied.

\subsubsection{Results for data set A} \label{sect221}

The re-analysis of data set A yielded a \gray{} excess of 277 counts
above the background (Table \ref{Tab1}), corresponding to a firm
detection with a statistical significance of 8.4$\sigma$ following
the method of \citet[][]{Li1983}. The increase in significance with
respect to the published result in \citet[][]{paper1} is related to
the application of improved analysis techniques.

\begin{table}
\caption{H.E.S.S. data and analysis results. The first column
represents the data set.
The second and third columns show the number of signal + background events
around the source position, and background events from the
off-source region, respectively. The fourth column shows the excess in
\grays{}.
The background normalisation ($\alpha$) is $\approx 0.022$.}
\label{Tab1}
\begin{tabular}{| c | c | c | c |}
\hline Data set name & On & Off & Excess  \\
 & (counts) & (counts) & (counts) \\
\hline
A & 1242 & 44308 & 277 \\
B &  928 & 30850 & 245 \\
Combined & 2170 & 75158 & 522 \\
\hline
\end{tabular}
\end{table}

We derive the energy spectrum using a forward-folding technique \citep[][]{Piron2001}.
The analysis threshold, $E_{\mathrm{thr}}$\,=\,0.25\,TeV, is given by the energy at which
the effective area falls to 20\% of its maximum value. The likelihood maximisation
for a power-law hypothesis, $dN/dE=N_{0} \times(E/E_{0})^{-\Gamma}$,
yields a photon index of $\Gamma=2.51\pm0.19_{\mathrm{stat}}\pm0.20_{\mathrm{sys}}$
and a normalisation constant of
$N_{0}=(1.44\pm0.22_{\mathrm{stat}}\substack{+0.43 \\
-0.29}_{\mathrm{sys}})\times10^{-13}$ cm$^{-2}$ s$^{-1}$ TeV$^{-1}$
at $E_0$=1 TeV. The main and cross-check analyses used in this paper
provide compatible results. This ImPACT analysis leads to a smaller
statistical error on the photon power-law index compared with the
previously published value,
$\Gamma=2.73\pm0.45_{\mathrm{stat}}\pm0.20_{\mathrm{sys}}$.
The central value of the normalisation coefficient
obtained with the ImPACT analysis is lower by a factor of 1.7 than
the previously reported value, but they are still marginally
compatible within statistical and systematic errors. Accumulation of
the exposure time of  data set B in addition to that of the data
set A allows us to refine the consistency between the current
results and the previously published results by redoing a
Hillas-based analysis with the latest calibration values (see
Appendix A for details). The accuracy of the calibration has
been considerably improved since 2009 and this in turn leads to a
minimisation of the systematic uncertainty on the flux normalisation
of faint VHE \gray{} sources with large exposure time, such as the
Cen A core. The systematic uncertainties are conservatively
estimated to be $\pm0.20$ on the photon index and $\substack{+30\% \\
-20\%}$ on the normalisation coefficient.

\subsubsection{Results for data set B}\label{sect222}

The analysis of data set B yielded a \gray{} excess of 245 counts
above the background (Table \ref{Tab1}). This \gray{} excess
corresponds to a firm detection with a statistical significance of
8.8$\sigma$. Thus, the Cen A \gray{} core is clearly detected as a
source of VHE emission in both of the H.E.S.S. data sets.
The spectral analysis of the data taken in 2009-2010 yields a
photon index of $\Gamma=2.55\pm0.19_{\mathrm{stat}}\pm0.20_{\mathrm{sys}}$ and a
normalisation constant of
$N_{0}=(1.50\pm0.22_{\mathrm{stat}}\substack{+0.45 \\ -0.30}_{\mathrm{sys}})\times10^{-13}$ cm$^{-2}$
s$^{-1}$ TeV$^{-1}$ at $E_0$=1 TeV.

To search for variability between the data sets A and B, one
needs to compare the intrinsic spectral properties of the source
in these two time intervals.
A comparison of the spectral analyses of the H.E.S.S. data sets A
and B shows that the values of the power-law photon indices are
compatible with each other and with that previously reported. As for
the normalisations of the VHE spectrum of the Cen A \gray{} core,
the best-fit normalisation values obtained with the analysis of both
the data sets are compatible with each other and somewhat lower than
(yet marginally compatible with) the previously reported value.

\subsubsection{Results for the combined H.E.S.S. data from 2004-2010}\label{sect223}

\begin{figure}
    \centering
    \includegraphics[width=0.5\textwidth]{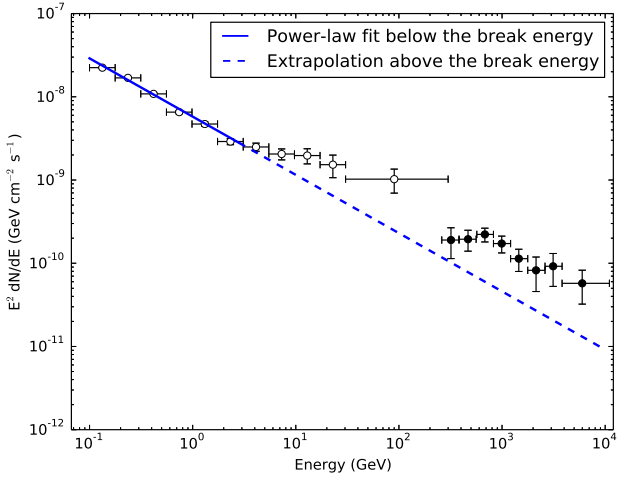}
    \caption{SED of Cen A \gray{} core.
    \textit{Fermi}-LAT and H.E.S.S. data points along with a high-energy
    power-law extrapolation of the $\gamma$-ray spectrum measured below the
    break energy. Eight years of \textit{Fermi}-LAT data and 213 hours of
    H.E.S.S. data were used. Statistical error bars are shown.}
    \label{FigSpec}
\end{figure}

Applying the ImPACT analysis to the combined data set, an excess of 522 events above the background
is detected (Table \ref{Tab1}). This excess leads to a
firm detection of the Cen A \gray{} core with H.E.S.S. at a
statistical significance of 12$\sigma$.
The same spectral analysis as before is applied to the full data set
and yields a photon index of $\Gamma=2.52
\pm0.13_{\mathrm{stat}}\pm0.20_{\mathrm{sys}}$ and a normalisation
constant of $N_{0}=(1.49\pm0.16_{\mathrm{stat}}\substack{+0.45 \\
-0.30}_{\mathrm{sys}})\times 10^{-13}$ cm$^{-2}$ s$^{-1}$ TeV$^{-1}$
at $E_0$=1 TeV. The reconstructed spectrum of the Cen A \gray{} core
is shown in Fig. \ref{FigSpec}. All of the eight SED data points in
the VHE range are above a 2.5$\sigma$ significance level, while only
one SED data point exceeds a 2$\sigma$ significance level in
\citet[][]{paper1}. The derived data points for each energy band in
the VHE range, shown in Fig. \ref{FigSpec}, agree within error bars
with those for the first and second data sets. The VHE spectrum of
the Cen A core is compatible with a power-law function ($\chi^2=3.9$
with 6 DOF).

If one takes the values of the spectral parameters from
the LAT four-year Point Source Catalogue (3FGL) \citep[][]{3fgl} obtained
from the \textit{Fermi}-LAT observations of Cen A between 100 MeV
and 100 GeV assuming a single power-law spectrum, then one finds
that $N_{0}=(0.45\pm0.07)\times10^{-13}$ ph
cm$^{-2}$s$^{-1}$TeV$^{-1}$ at 1 TeV and $\Gamma=2.70\pm0.03$.
Therefore, the differential flux at 1 TeV derived from the H.E.S.S.
observations in 2004-2010 is about 3.5 times larger than that
inferred from a power-law extrapolation of the 3FGL catalogue
spectrum. This indicates that a deviation of the spectrum from a
single power law (``hardening'') should occur at GeV energies to
match the TeV data (see Section 3).

We searched the combined data set for evidence of time variability
at the position of the Cen A core. No significant variability was
found on timescales of 28 minutes (individual runs), months, or
years. The lack of apparent flux variability along with no change in
spectral parameters between the two data sets justifies combining
all available data when comparing the spectrum to that of
\textit{Fermi}-LAT. We note that given the low flux level of the Cen
A \gray{} core, a flux increase by a factor of approximately ten
would be needed to allow a significant detection of variability on timescales
of 28 minutes (corresponding to a $5\sigma$ detection in individual
runs).

\section{\textit{Fermi}-LAT observations and results}
In HE \grays{}, the core of Cen~A is firmly detected with the
\textit{Fermi}-LAT using eight years of Pass~8 data spanning over three
orders of magnitude in energy. LAT analysis of Cen~A involves unique
challenges not present in other individual extragalactic source
analyses, largely due to the massive angular extent of the Cen~A
non-thermal outer lobes of $\sim9^{\circ}$ and the proximity to the
Galactic plane (Galactic latitude $\approx19.4^{\circ}$), which is a
bright source of diffuse $\gamma$-ray emission. In the following, we
report corroborating evidence for the presence of an additional
spectral component at $\gamma$-ray energies above a break of
$\simeq2.8$ GeV. No significant variability either above or below
this break has been detected.

\subsection{Observations and analysis}

LAT is a pair-conversion telescope on the FGST \citep[][]{atwood09}.
It has a large field of view ($\sim$2.4 sr) and has been scanning
the entire sky continuously since August 2008. The broad energy
coverage and the all-sky monitoring capability make LAT
observations, which bridge the gap between soft $\gamma$-ray (MeV)
and TeV energy ranges, crucial to explore the spectrum of the Cen~A
high-energy core and to test its variability.

We selected Pass 8 \texttt{SOURCE} class \textit{Fermi}-LAT photon
data spanning eight years between August 4, 2008 and July 6, 2016 (MET
239557417 to 489507985) with energies between 100\,MeV and 300\,GeV.
Higher energies than 300\,GeV yield no detection. We performed a
binned analysis by choosing a $10^{\circ}\times10^{\circ}$ square
region of interest (ROI) centred at the position of the Cen~A core
(3FGL J1325.4-4301) as reported in the 3FGL catalogue, R.A.
$=201\fdg367$, Decl. $=-43\fdg030$ \citep{3fgl}, with spatial bins
$0\fdg1$ in size and initially eight energy bins per decade. We applied
standard quality cuts \texttt{(DATA\_QUAL==1 \&\& LAT\_CONFIG==1)}
and removed all events with zenith angle $>90^{\circ}$ to avoid
contamination from the Earth's limb. In the following, models are
compared based on the maximum value of the logarithm of the
likelihood function, $\mathrm{log}\mathcal{L}$. The significance of
model components or additional parameters is evaluated using the
test statistic, whose expression is
$\mathrm{TS}=2(\mathrm{log}\mathcal{L}-\mathrm{log}\mathcal{L}_0)$,
where $\mathcal{L}_0$ is the likelihood of the reference model
without the additional parameter or component \citep[][]{mattox96}.

To model the sources within the ROI, we began with sources from the
3FGL within the $15^{\circ}\times15^{\circ}$ region enclosing the
ROI \citep[the 3FGL models the Cen~A lobes with a template
created from $22$\,GHz \textit{WMAP} data; see][]{wmap}. We included
the isotropic and Galactic diffuse backgrounds,
\texttt{iso\_P8R2\_SOURCE\_V6\_v06} and \texttt{gll\_iem\_v06}
\citep{lat-iem}, respectively. We fixed the normalisations of both
the isotropic and Galactic diffuse source models to one to avoid
leakage of photons from the Cen A lobes into these templates; when
free, they converged to unrealistic values. The convergence to
unrealistic values is due to unmodelled emission from the Cen A
lobes. We introduced additional background sources in order to
account for excess lobe emission. After creation of the fully
developed model, freeing both these diffuse sources has a negligible
effect on the results. We optimised each source in the model
individually\footnote{Because of the large number of free parameters
due to the number of sources, we loop over all model components and
fit their normalisations and spectral shape parameters while fixing
the rest of the model so that the whole model converges closer to an
overall maximum likelihood.}, and then left the normalisation
parameters of sources within 3$^{\circ}$ and the spectral shapes of
only the core and lobes free during the final likelihood
maximisation. We generated a residual TS
map and residual
significance map for the ROI and found several regions with data
counts in excess of the model. A TS map is created by moving a
putative point source through a grid of locations on the sky and
maximising $\mathrm{log}\mathcal{L}$ at each grid point, with the
other, stronger, and presumably well-identified sources included in
each fit. New, fainter sources may then be identified at local
maxima of the TS map. Using the residual TS map as a guide for
missing emission, we added ten additional background sources to the
ROI model. These ten sources are most likely a surrogate for excess
lobe emission and should not be considered new individual point
sources. After re-optimisation and creation of a residual TS map, we
observe no significant ($>5\sigma$) regions of excess counts, and a
histogram of the residuals is well fit as a Gaussian distribution
centred around zero.

The precise \gray{} morphology of the Cen~A lobes is beyond the
scope of this work and is not needed to accurately determine the SED
of the core. This work on the Cen~A core does not require a
high-precision model for the lobes, as the angular size of the Cen~A
lobes is sufficiently larger than the point spread function (PSF) of
the LAT, especially at higher energies where this study is focused
($<1^{\circ}$ $95\%$ containment angle above $5$\,GeV)
\footnote{\url{https://www.slac.stanford.edu/exp/glast/groups/canda/lat\_Performance.htm}.}.
However, to verify this, we tested the modelling procedure above
using two alternative $\gamma$-ray templates of the Cen~A lobes. The
first of these was a modification to the public \textit{WMAP}
template involving ``filling in'' the $2^{\circ}$ diameter hole
surrounding the core. This was accomplished by patching this area
with nearby matching intensities. The second alternate lobe template
tested was one made from radio data from the Parkes telescope at
$6.3$\,cm wavelength \citep{parkes}. Use of these alternate lobe
templates had no significant effect on the resulting best-fit core
break energy or the flux above the break energy. However, we did
observe a flux deviation below the break energy, resulting in a drop
in the full band energy flux of the core by up to 17$\%$ depending
on which lobe template was being used. We believe this drop results
from the lack of a hole (circle containing values of 0) around the
core of the lobe template with the modified \textit{WMAP} and the
Parkes templates. We also introduced a version of the Parkes
template with a hole matching the one in the \textit{WMAP} and
observed a flux increase of 2$\%$ instead of a drop, lending
credence to our belief that the existence of the hole is the most
important factor for this analysis.

\subsection{Results of the observations of Cen~A with \textit{Fermi}-LAT}
We calculated an SED over the full range by dividing the data into
14 equally spaced logarithmic energy bins and then merging the four
highest energy bins into one for sufficient statistics. In each bin,
the Cen~A lobe and core spectral parameters were left free to
optimise and within each bin these spectra were fit using a single
power law. The resulting SED is plotted in Figs.~\ref{FigSpec} and
\ref{Fig4}. To plot the data point within the wide merged
energy bin, we used the prescription from \citet[][]{Lafferty95}.
The spectral hardening in the HE $\gamma$-ray emission from the core
of Cen~A above an energy break of 2.8 GeV is illustrated in
Fig.~\ref{FigSpec}. A broken power-law model describes well the
shape of the {\it{Fermi}}-LAT \gray{} spectrum with a break energy
of 2.8 GeV.

\begin{figure}
    \centering
    \includegraphics[width=0.5\textwidth]{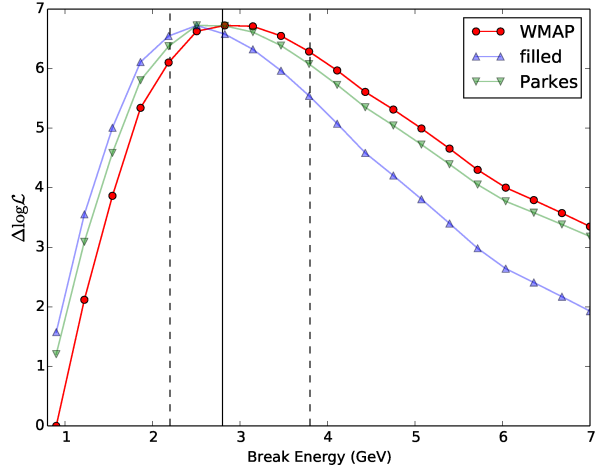}
    \caption{Change in overall $\mathrm{log}\mathcal{L}$ while fixing Cen~A core
    break energy to values within the range 0.9 -- 7.0\,GeV, as derived from
    \textit{Fermi}-LAT data using the \textit{WMAP}, modified \textit{WMAP}, and
    Parkes templates and compared to the $\mathrm{log}\mathcal{L}$ value at 0.9\,GeV
    for the \textit{WMAP} template. The solid vertical line shows
    the best-fit value of the break energy parameter, while the dashed vertical lines show
   $1\sigma$ interval for the parameter.}
    \label{breakprof}
\end{figure}

We optimised the break energy via a likelihood profile method. For
this purpose, we fixed all parameters in the ROI model except the
normalisations of sources within a 3$^{\circ}$ radius of the core to
their best-fit values from the full optimisation. The
$\mathrm{log}\mathcal{L}$ profiles for the broken power-law spectral
model and computed using the \textit{WMAP}, modified \textit{WMAP}, and Parkes templates
are plotted in Fig.~\ref{breakprof}.
From the position of the peak in the profile corresponding to
the \textit{WMAP} template, we find a best-fit break energy of $2.8\substack{+1.0 \\
-0.6}$\,GeV. To determine the statistical preference of the broken
power-law model over the single power law, we subtract the overall
$\mathrm{log}\mathcal{L}$ from the same ROI model with a single
power law from the $\mathrm{log}\mathcal{L}$ from the break energy
profile at $2.8$\,GeV. Because these models are nested, Wilks'
theorem yields a preference of the broken power law with $4.0\sigma$
confidence ($\chi^2=19.0$ with 2 DOF).

From this fully optimised $\gamma$-ray model of the Cen~A core, we
obtain a strong detection at 73$\sigma$ statistical level and
calculate a full-band energy flux
of $(4.59\pm0.14_{\mathrm{stat}} \substack{+0.17 \\
-0.13}_{\mathrm{sys, A_{eff}}})\times10^{-5}$\,MeV cm$^{-2}$
s$^{-1}$. The best-fit broken power-law
prefactor\footnote{\burl{https://fermi.gsfc.nasa.gov/ssc/data/analysis/scitools/source\_models.html\#BrokenPowerLaw}.}
is $(3.64\pm0.15)\times10^{-13}$\,cm$^{-2}$ s$^{-1}$ MeV$^{-1}$. In
the lower-energy band, we find a photon
index of $2.70\pm0.02_{\mathrm{stat}} \substack{+0.05 \\
-0.03}_{\mathrm{sys, A_{eff}}}$, and in the
higher band, $2.31\pm0.07_{\mathrm{stat}}\substack{+0.01 \\
-0.04}_{\mathrm{sys, A_{eff}}}$. This provides corroborating
evidence for a spectral hardening by $\Delta\Gamma\sim0.4$ above the
break energy.
Comparisons of these results to the Cen A
core spectrum from the 3FGL catalogue \citep[][]{3fgl} are not
meaningful, since their analysis did not include modelling of the
Cen A core spectrum as a broken power law. Using the modified
\textit{WMAP} template we observe a consistent photon index in the
lower and upper bands, respectively, of $2.68\pm0.03$ and
$2.26\pm0.07$, and using the Parkes template, $2.67\pm0.03$ and
$2.29\pm0.07$. We also tested for a log-parabola spectral shape
using a likelihood ratio test, analogous to \texttt{Signif\_Curve}
in the 3FGL catalogue, which \citet[][]{3fgl}
calculated
as 2.3\,$\sigma$, and found a TS$_{\mathrm{curve}}=4.5$, or
$\sim2.1$\,$\sigma$. The power-law index that we observe above the
spectral break is consistent with the index above 10\,GeV found in
the 3FHL catalogue \citep[][]{3fhl}.

Finally, we tested for variability of the Cen~A core both above and
below the break energy ($2.8$\,GeV) by calculating light curves
using a single power-law spectral model for each. Below the break,
we divided the data into 64 45-day bins and calculated flux
variability using the method described in \cite{2fgl} Sect. 3.6, with
systematic correction factor $f=0.02$. Keeping the power-law index
fixed to $2.70$, we calculate $0.09\sigma$ ($\chi^2=47.3$ with 63
DOF) significance for flux variability. Above the break, we divided
the data into nine-month bins. Keeping the power-law index fixed to
$2.31$, we do not see evidence for flux variability ($1.9\sigma$,
$\chi^2=16.6$ with 9 DOF).

\section{Discussion}

\subsection{Beyond a single-zone SSC description of the $\gamma$-ray
core SED of Cen A}
The proximity and the diversity of the radio structures associated
with the activity of its core make Cen A an ideal laboratory to
investigate radiative processes and jet physics. In this regard, an
improved characterisation of its SED is important in distinguishing
which emission component is likely to dominate the observed
radiation. Earlier investigations \citep[e.g.][]{chiaberge01}
suggested that the SED of the core of Cen A  (i.e. the central
source unresolved with radio, infrared, hard X-ray, and \gray{}
instruments) up to sub-GeV energies appears remarkably similar to
that of blazars. In a $\nu$-$\nu F_{\nu}$ plot, the SED seems well
represented by two broad peaks, one located in the far-infrared band
and the other in the \gray{} band at energies $\sim0.1$ MeV. The SED
as known prior to 2009 was satisfactorily described by a single
zone, homogeneous SSC model assuming the jet to be misaligned (i.e.
lower Doppler boosting compared to blazars). The detection of VHE
and HE $\gamma$ rays from Cen A by H.E.S.S. and \textit{Fermi}-LAT
has started to complicate this simple picture. If the available
(non-contemporaneous) H.E.S.S. and \textit{Fermi}-LAT data are
added, a single zone SSC model is no longer able to adequately
account for the overall core SED of Cen~A \citep[][see also
Roustazadeh \& B\"ottcher~2011, Petropoulou et al. 2014]{abdo10}.
The SSC spectral component introduced earlier
\citep[][]{chiaberge01} appears to work well only for the radio band
to the MeV \gray{} band.

Moreover, the detection of VHE $\gamma$ rays compatible with a
power law up to $\sim5$ TeV raises the principal challenge of
avoiding internal (i.e. on co-spatially produced synchrotron
photons) $\gamma\gamma$ absorption in a one-zone SSC approach.
Interferometric observations with the MID-infrared Interferometeric
instrument (MIDI) at the Very Large Telescope Interferometer
array \citep[][]{meisenheimer07}
showed that the mid-infrared (MIR) emission from the core of Cen A
is dominated by an unresolved point source $<10$ mas (or $<0.2$ pc).
\citet[][]{abdo10} have argued that the MIR and VHE emission cannot
originate in the same region, since the VHE emission would be
strongly attenuated due to $\gamma\gamma$ interaction with
mid-infrared (soft) photons. The strength of this argument depends
on how well possible Doppler boosting effects can be constrained,
that is, on inferences with respect to the inclination and the bulk flow
Lorentz factor  of the sub-parsec
scale jet in Cen~A.
It could be shown
by extending the argumentation from Section 5.2 of \citet[]{abdo10}
that the $\gamma\gamma$-attenuation problem might be alleviated if
the sub-parsec
jet were inclined at 11$^{\circ}$, that is,
slightly below the lower limit of the angular range
$\theta\sim12^{\circ}-45^{\circ}$ allowed by recent Tracking Active
Galactic Nuclei with Austral Milliarcsecond Interferometry (TANAMI)
monitoring constraints on the sub-parsec
scale jet \citep[][]{muller14}.
Motions with the Doppler factors required to avoid $\gamma\gamma$
attenuation ($\delta_{\mathrm{D}}>5.3$), however, have not yet been
observed on sub-parsec scales.

The previously mentioned considerations, along with the evidence for
a clear hardening of the HE spectrum of Cen~A, make a single-zone SSC
interpretation for its overall SED very unlikely. Alternative
scenarios, where the TeV emission from the high energy Cen A core is
associated with the presence of an additional emission component is
instead favoured.

\subsection{Characterising the overall core SED with other multi-wavelength observations}

\begin{figure}
\centering
\begin{minipage}{0.47\textwidth}
\includegraphics[angle=0,width=8.7cm]{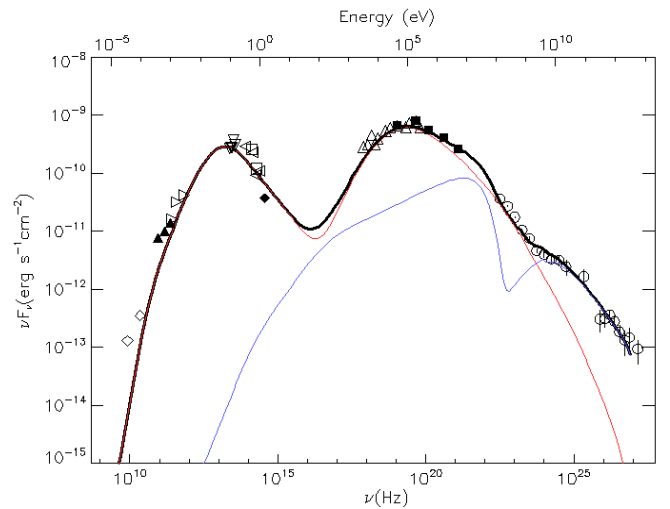}
\caption{SED of  Cen A core with model fits as described in text. The red curve corresponds to an SSC component
designed to fit the radio to sub-GeV data. The blue curve
corresponds to a second SSC component added to account
for the highest energy data. The black curve corresponds to the sum
of the two components. SED points as derived from H.E.S.S. and
\textit{Fermi}-LAT data in this paper are shown with open circles.}
\footnotetext{Observations from the radio band to the MeV \gray{}
band are from TANAMI ($\diamond$), SEST ($\blacktriangle$), JCMT
($\triangleright$), MIDI ($\triangledown$), NAOS/CONICA
($\triangleleft$), NICMOS ($\square$), WFPC2 ($\blacklozenge$),
{\it{Suzaku}} ($\triangle$), OSSE/COMPTEL ($\blacksquare$).
The acronyms are described in Appendix B.} \label{Fig4}
\end{minipage}
\end{figure}

A variety of multi-wavelength data, albeit with varying angular
resolution and taken non-contemporaneously, is available for Cen~A
and can be used to construct a characteristic core SED, an example
of which is presented in Fig. \ref{Fig4}. Observations in different
broad energy ranges are shown with different symbols.
In the $\gamma$-ray regime, we combine H.E.S.S.
and \textit{Fermi}-LAT data
to build a quasi-contemporaneous
high-energy core SED. One should keep in mind, however,
that given the angular resolution
of H.E.S.S. and \textit{Fermi}-LAT, the large-scale jet and inner
lobes of Cen~A could in principle also contribute to the observed
\gray{} signal.

Cen A is the highest flux radio galaxy detected in hard X-ray and
MeV \gray{} bands. As can be seen from Fig.~\ref{Fig4}, this energy
range plays an important role in the modelling of its emission. The
angular resolution at these energy bands is relatively poor compared
to that at other energies (including radio, infrared, soft X-rays,
GeV, and VHE
 \grays{}). It corresponds to about $2\fdg{5}$
for INTEGRAL SPI in the bandpass 18 keV-8 MeV and to about
4$^{\circ}$ in the energy range 1-30 MeV for COMPTEL
\citep[][]{steinle98, Steinle10}. We note that a recent spectral
analysis of ten years of observations with INTEGRAL SPI favours a
jet origin for the hard X-ray emission \citep[][]{burke14},
supporting the proposal that the second peak in the SED of the Cen A
core (with a maximum at $\sim0.1$ MeV) is jet-related and probably
due to SSC radiation \citep[][]{chiaberge01,abdo10}. A possible
X-ray contribution from accretion, however, cannot yet be excluded
\citep[for discussion, see][]{evans04, meisenheimer07, furst16}. The
available archival data measured in hard X-rays and MeV \grays{} of
Cen A have been included in Fig.~\ref{Fig4}.
For the lower-energy SED part, which includes radio, mm-, infrared
and optical data points, and seems well described by a
synchrotron source, the available archival data are taken from
\citet[][]{meisenheimer07}, with the exception of two data points at
8.4 GHz and 22.3 GHz measured on 2009 November 27 and 29 as part of
the TANAMI programme \citep[][]{ojha10}, replacing three consistent
radio data points that were measured in the mid-1990s.

\subsection{Modelling the high-energy core SED with a second emission component}

The observed smooth TeV spectrum and the spectral hardening by
$\Delta\Gamma\sim0.4$ as observed with H.E.S.S. and
\textit{Fermi}-LAT are strongly suggestive of the contribution of a
second emission component in addition to the conventionally employed
single-zone SSC component under the assumption of a misaligned jet.
A variety of different (not mutually exclusive) scenarios for the
physical origin of this second emission component could be
envisaged. Proposals in the literature for Cen~A encompass

(a) magnetospheric (pulsar-like) scenarios based on leptonic inverse
Compton processes in a radiatively inefficient disk environment
\citep{rieger09,rieger11},

(b) inner (parsec- and sub-parsec-scale) jet models involving for example
multiple leptonic SSC-emitting components travelling at different
angles to the line of sight \citep{lenain08}, inverse Compton
interplay in a stratified jet geometry \citep{ghisellini05},
photo-meson $p\gamma$-interactions of ultra-high-energy protons in
strong (e.g. standard disk-type) photon fields
\citep{kachelriess10,sahu12,petropoulou14,fraija2014} and elaborated
lepto-hadronic modifications thereof \citep{reynoso11, cerruti16},
or $\gamma$-ray-induced pair-cascades in a strong accretion disk
field \citep{sitarek10}, a dusty torus-like region
\citep{roustazadeh2011}, or a starlight photon field
\citep{stawarz06},

(c) extended astrophysical scenarios involving for example hadronic
$pp$-interactions of accelerated protons with ambient matter in its
kiloparsec-scale region \citep{sahakyan13}, the combined high-energy
$\gamma$-ray contribution from a supposed population of millisecond
pulsars \citep{brown16}, or leptonic inverse-Compton scattering off
various photon fields (SSC, host galaxy starlight, cosmic microwave
background, extragalactic background light) in the kiloparsec-scale
jet of Cen~A \citep{stawarz03, hardcastle11}, and

(d) explanations involving physics beyond that of the Standard
Model, for example the self-annihilation of dark matter (DM) particles in a
putative central dark matter spike \citep{brown16}.

Some critical astrophysical questions arise in each of these models:
near-black-hole scenarios, for example, require advection-dominated
accretion disk environments to satisfy external opacity conditions,
leptonic models often deviate significantly from equipartition and
are affected by internal opacity constraints, hadronic scenarios
usually require a very high jet power, and pulsar-population models
are dependent on poorly-known density profiles, while DM models need
anomalously high dark matter concentrations. However, the limited
angular resolution of current \gray{} instruments and the fact that
no significant statistical evidence for variability of the \gray{}
emission above the break (neither at \textit{Fermi}-LAT nor VHE
energies) has been found, does not make it possible to strongly
exclude any of these models. We note, though, that any (future) hint
of variability would likely disfavour models of type (c)-(d).
The (apparent) lack of variability, on the other hand, could simply be
a matter of limited statistics and therefore might still be reconciled
with inner jet-related scenarios.
The increased sensitivity of the Cherenkov Telescope Array
\citep[CTA;][]{cta17} will enable a deeper probe into this and may
eventually distinguish between models and resolve the physical
nature of this component.

Noting these limitations, we nevertheless would like to provide an
illustration here that the current core SED could be satisfactorily
modelled by two jet-related components where the emission below the
break is attributed to the conventional (misaligned) SSC-emitting
component and the emission above the break to an additional
SSC-emitting jet component. We model both components as jet blobs of
different size and magnetic field strength. Assuming that the
conventional single-zone SSC description works well for the radio to
sub-GeV part of the spectrum, we adopt the same parameters (see
Table~\ref{Tab3}) for the first SSC component as reported
earlier\footnote{The power-law index, $p_1$, of 1.8 for the first
component was adopted from \citet[][]{abdo10} in order to fit the
exceptionally flat Cen A spectrum,
$F_{\mathrm{\nu}}\propto\nu^{-0.36}$, between $10^{11}$ and
$3\times10^{13}$ Hz \citep[][]{meisenheimer07}.} \citep[][]{abdo10},
apart from considering a self-consistent maximum electron Lorentz
factor of $\gamma_{\mathrm{max}} =10^7$. The SED is modelled using
the numerical code SED
Builder\footnote{\burl{https://tools.asdc.asi.it/}.}
\citep[][]{massaro06, tramacere09, tramacere11}. To account for the
$\gamma$-ray spectrum above the break, we introduce a second
SSC-emitting zone for which we require, amongst others, that (a) the
energy density in the particles is comparable to (or less than) the
energy density in the magnetic field $B^2/(8\pi)$ (one-sided
equipartition constraint), (b) the
dynamical timescale $\approx R/c$ is larger than the synchrotron
cooling timescale at high energies (efficiency constraint), (c) the
synchrotron loss timescale is longer than the gyro-timescale at
$\gamma_{\mathrm{max}}$ (acceleration constraint), and (d) the
optical depth to internal $\gamma\gamma$ absorption is less than one
(opacity constraint). The model given for the second SSC component
(Table \ref{Tab3}) provides an exemplary set of parameters that
satisfy these constraints and that satisfactorily reproduces the
observed spectrum. While not unique, this example provides an
illustration that both the VHE emission measured with H.E.S.S. as
well as the GeV emission measured with \textit{Fermi}-LAT could be
accounted for by means of a two-zone SSC scenario. If one relaxes
the requirements (e.g. the one-sided equipartition constraint above),
additional descriptions with for example a rather low magnetic field
strength, become possible \citep[e.g.][]{abdo10}. More complex
realisations might perhaps be possible if the second component were characterised by a different (blazar-like) Doppler factor
$\delta_\mathrm{D}>1$.

\begin{table*}
\centering \caption{Parameters used for modelling overall core
SED of Cen A with two SSC-emitting components.}
\begin{tabular}{ | c | c | c | c |}
\hline
Parameter & Symbol & the 1st SSC zone & the 2nd SSC zone\\
\hline
Doppler factor & $\delta_{\mathrm{D}}$ & 1.0 & 1.0 \\
Jet angle & $\theta$ & 30$^{\circ}$ & 30$^{\circ}$\\
Magnetic field (G) & $B$ & 6.2 & 17.0\\
Comoving blob size (cm) & $R_{\mathrm{b}}$ & $3.0\times10^{15}$ & $8.8\times10^{13}$\\
\hline
Low-energy electron power-law index & $p_{1}$ & 1.8 & 1.5 \\
High-energy electron power-law index & $p_{2}$ & 4.3 & 2.5 \\
Minimum electron Lorentz factor & $\gamma_{\mathrm{min}}$ & $3\times10^{2}$ & $1.5\times10^{3}$\\
Maximum electron Lorentz factor  & $\gamma_{\mathrm{max}}$ & $1\times10^{7}$ & $1\times10^{7}$\\
Break electron Lorentz factor & $\gamma_{\mathrm{brk}}$ & $8.0\times10^{2}$ & $3.2\times10^{4}$\\
Electron energy density (erg cm$^{-3}$) & $\epsilon$ & $1.3$ & $7.8$ \\
\hline
\end{tabular}
\label{Tab3}
\end{table*}

\section{Conclusions}
High-energy observations of the core region in active galaxies provide important insights into
the physical processes driven by a central powerhouse containing an accreting, jet-emitting
supermassive black hole system.
In the case of Cen~A, the H.E.S.S. discovery of VHE \gray{} emission from its central
region \citep[][]{paper1} exceeded expectations from conventional (mis-aligned) single-zone
SSC scenarios, casting doubt on the appropriateness of such an interpretation.
Non-simultaneous \textit{Fermi}-LAT results \citep[][]{sahakyan13, brown16} are indeed
indicative of a transition region above a few GeV in the $\gamma$-ray core spectrum of
Cen~A and provide evidence that the VHE emission is associated with an additional
radiative component.

This paper reports results of new (more than 100 hr) VHE
observations of the Cen~A $\gamma$-core with H.E.S.S. accumulated
during the {\it{Fermi}}-LAT operation and provides a detailed
characterisation of the complete VHE data set using advanced
analysis methods. VHE $\gamma$-ray emission from the core of Cen~A
is detected at 12$\sigma$. No significant variability is apparent in
the VHE data set. A spectral analysis of the complete data set
yields a photon index of $\Gamma=2.52 \pm
0.13_{\mathrm{stat}}\pm0.20_{\mathrm{sys}}$ and a normalisation
constant of $N_{0}=(1.49\pm0.16_{\mathrm{stat}}\substack{+0.45 \\
-0.30}_{\mathrm{sys}})\times 10^{-13}$ cm$^{-2}$ s$^{-1}$ TeV$^{-1}$
at $E_0$=1 TeV. Spectral analyses of the H.E.S.S. data taken before
and after the launch of the \textit{Fermi} satellite give comparable
results and validate the construction of a joined \gray{} spectrum
based on {\it Fermi}-LAT and H.E.S.S. data. We also present an
update of the Cen~A core spectrum at GeV energies using eight years of
{\it Fermi}-LAT Pass 8 data. The \textit{Fermi}-LAT analysis
provides clear evidence at a level of 4.0 $\sigma$ for spectral
hardening by $\Delta\Gamma \simeq 0.4$ at $\gamma$-ray energies
above a break energy of $\simeq2.8$\,GeV. This hardening is
suggestive of an additional $\gamma$-ray emitting component
connecting the HE emission above the break energy to the one
observed at VHE energies. Both the hardening of the spectrum above
the break energy at a few GeV and the VHE emission excess over the
power-law extrapolation of the \gray{} spectrum measured below the
break energy are a unique case amongst the VHE AGNs. The results
allow us for the first time to construct a representative
(contemporaneous) HE-VHE SED for Cen~A. While a variety of different
interpretations are available, the physical origin of the additional
\gray{} emitting component cannot yet be resolved due to
instrumental limitations in angular resolution and the apparent
absence of significant variability in both the HE and VHE data. It
is possible, however, that the additional emission component is
jet-related and we provide one SSC model fit to illustrate this.

Despite their faintness at \gray{} energies,
radio galaxies such as Cen~A are emerging as a unique \gray{} source
population offering
important physical insight beyond what could usually be achieved in
classical blazar sources. With its increased sensitivity, CTA is
expected to probe deeper into this and help to eventually resolve
the nature of the \gray{} emission in Cen~A.

\begin{acknowledgements}

The support of the Namibian authorities and of the University of
Namibia in facilitating the construction and operation of H.E.S.S.
is gratefully acknowledged, as is the support by the German Ministry
for Education and Research (BMBF), the Max Planck Society, the
German Research Foundation (DFG), the Helmholtz Association, the
Alexander von Humboldt Foundation, the French Ministry of Higher
Education, Research and Innovation, the Centre National de la
Recherche Scientifique (CNRS/IN2P3 and CNRS/INSU), the Commissariat
\`{a} l'\'{e}nergie atomique et aux \'{e}nergies alternatives (CEA),
the U.K. Science and Technology Facilities Council (STFC), the Knut
and Alice Wallenberg Foundation, the National Science Centre, Poland
grant no. 2016/22/M/ST9/00382, the South African Department of
Science and Technology and National Research Foundation, the
University of Namibia, the National Commission on Research, Science
\& Technology of Namibia (NCRST), the Austrian Federal Ministry of
Education, Science and Research and the Austrian Science Fund (FWF),
the Australian Research Council (ARC), the Japan Society for the
Promotion of Science, and by the University of Amsterdam. We
appreciate the excellent work of the technical support staff in
Berlin, Zeuthen, Heidelberg, Palaiseau, Paris, Saclay, T\"{u}bingen,
and in Namibia in the construction and operation of the equipment.
This work benefited from services provided by the H.E.S.S. Virtual
Organisation, supported by the national resource providers of the
EGI Federation.

\textit{Fermi} LAT Collaboration acknowledges generous ongoing support
from a number of agencies and institutes that have supported both the
development and the operation of the LAT as well as scientific data analysis.
These include the National Aeronautics and Space Administration and the
Department of Energy in the United States, the Commissariat \`a l'Energie
Atomique and the Centre National de la Recherche Scientifique / Institut
National de Physique Nucl\'eaire et de Physique des Particules in France,
the Agenzia Spaziale Italiana and the Istituto Nazionale di Fisica Nucleare
in Italy, the Ministry of Education, Culture, Sports, Science and Technology
(MEXT), High Energy Accelerator Researc Organization (KEK) and Japan
Aerospace Exploration Agency (JAXA) in Japan, and the K.~A.~Wallenberg
Foundation, the Swedish Research Council and the Swedish National Space
Board in Sweden.

Additional support for science analysis during the operations phase is
gratefully acknowledged from the Istituto Nazionale di Astrofisica
in Italy and the Centre National d'\'Etudes Spatiales in France.
This work performed in part under DOE Contract DE-AC02-76SF00515.

\end{acknowledgements}

\begin{appendix}
\section{Comparison of spectral results from Hillas-based analyses}

To perform a Hillas-based analysis, we applied a standard cut of
$\theta^2 < 0.0125$ deg$^2$ for the calculation of the number of ON
events, where $\theta^2$ is the square of the angular separation
between the reconstructed shower position and the source position.
This cut is optimised to minimise the contamination by the
background and is somewhat different to that used in the previous
publication ($\theta^2 <0.03$ deg$^2$). The Hillas-based analysis
performed here for the combined data set (A+B) results in a lower
value of the normalisation coefficient compared with the published
value. The obtained value is compatible with those derived with the
main (ImPACT) and cross-check analyses. The compatibility of these
results gives us confidence in the reliability of the current
cross-checked analysis of the Cen A core. The comparison suggests a
wider range of the systematic errors for the results of the
Hillas-based analysis of data set A than that estimated
in \citet[][]{paper1}. The lack of temporal variability in flux
between data sets A and A+B concluded from the ImPACT analyses
supports this suggestion.

\section{List of acronyms}

SEST - 15 m Swedish-ESO Submillimetre Telescope; JCMT - 15 m James
Clerk Maxwell Telescope;
NAOS/CONICA - Nasmyth Adaptive Optics System/Coude Near Infrared
Camera; NICMOS - Near Infrared Camera and Multi-Object Spectrometer;
WFPC2 -  Wide Field and Planetary Camera 2; Swift-BAT - Swift-Burst
Alert Telescope;
BATSE - Burst and Transient Source Experiment;
OSSE - Oriented Scintillation Spectrometer
Experiment; COMPTEL - imaging COMPton TELescope;
EGRET - Energetic Gamma Ray Experiment Telescope.

\end{appendix}

\bibliography{refsVersion6}

\newpage

\end{document}